\documentclass[aps,prb,twocolumn,showpacs,superscriptaddress,groupedaddress]{revtex4-1}
\usepackage{graphicx}
\usepackage{dcolumn}
\usepackage{bm}
\usepackage{amssymb}
\usepackage{subfigure}
\usepackage{color}

\begin{document}
\title{A Quantum Approach of Meso-Magnet Dynamics with Spin Transfer Torque}
\author{Yong Wang}\altaffiliation{Present address: Department of Physics, University of Hong Kong}
\author{L. J. Sham}
\email{lsham@ucsd.edu}
\affiliation{Center for Advanced Nanoscience, Department of Physics,
University of California, San Diego, La Jolla, California 92093-0319, USA}
\begin{abstract}
We present a theory of magnetization dynamics driven by spin-polarized current  in terms of the quantum master equation. In the  spin coherent state representation, the master equation becomes a Fokker-Planck equation, which naturally includes the spin transfer and quantum fluctuation.  The current electron scattering state is correlated to the magnet quantum states, giving rise to quantum correction to the electron transport properties in the usual semiclassical theory. In the large spin limit, the magnetization dynamics is shown to obey the Hamilton-Jacobi equation or the Hamiltonian canonical equations. 
\end{abstract}
\pacs{75.78.-n, 05.10.Gg, 85.75.-d}
\maketitle
\section{Introduction}
The theory of open quantum systems, which has been greatly developed in the past several decades, plays a critical role in the understanding and control of the dynamics of quantum systems that are coupled to the surrounding environments.\cite{OpenQuan1,OpenQuan1a,OpenQuan2} The basic issues such as  dissipation,  decoherence,  measurement, and noise source, etc. of the systems are usually investigated in the open quantum system framework.   This theory has  wide applications in the fields including quantum optics,\cite{OpenQuan1,OpenQuan1a,OpenQuan2} ultracold atoms,\cite{arecchi72} and quantum information and  computation.\cite{weedbrook12} On the other hand, in micromagnetics\cite{aharoni96,suhl07} and spintronics\cite{revtorq1,revtorq2} the magnetization dynamics is commonly treated as classical even though the control and dissipation parameters are couched as from quantum sources. While these methods have been highly successful in simulating magnetization reversal and spin-torque driven magnetization dynamics,\cite{LLG} there is the question whether there are quantum effects not exhibited by these classical treatments, when the magnet is in the mesoscopic range, of $10^3 - 10^7$ spins, between the molecular magnets and the macroscopic magnets as defined by Ref.~\onlinecite{wernsdorfer07}.
Addressing this question may not only pave the way for the future technology developments, but also broaden our vision and deepen our understanding of the emerging mesoscopic quantum world between the well established microscopic and macroscopic ones. In this paper, we present a theory of a single domain mesoscopic magnet as a member of the family of open quantum systems for its spin-current-driven dynamics. 

When the spin-polarized current passes through the ferromagnetism (FM) layer, the spin angular momentum of the current electrons is transferred to the FM layer and thus rotate the magnetization\cite{torq1,torq2}. This so-called ``spin transfer torque (STT)" effect has now become the most important method to control the magnetization dynamics in the nano-scale structures, where the conventional Oersted field generated by the electric current is less practical\cite{revtorq1,revtorq2}. Numerous magnetoelectronics devices have been proposed and fabricated based on the STT-driven magnetization dynamics.\cite{revtorq1,revtorq2} In spite of its great success and importance, the fundamental physics of STT in the standard theory is semiclassical. The magnetization dynamics is described by the classical Landau-Lifshitz-Gilbert (LLG) equation\cite{LLG}, while the STT terms in the LLG equation is obtained from the quantum scattering of the spin current electrons by the classical potential of the magnetization\cite{torq1,torq2}. This semiclassical picture is expected to break down as the magnet is further miniaturized to the mesoscopic regime. Furthermore, the STT has been used to stimulate and control the spin waves\cite{Magnon}, or magnons. Therefore, a more sophisticated investigation of the STT in the full quantum picture is necessary in order to adapt the new developments in the field. In a previous study, we shown that the continuous scatterings between the quantum macrospin state of a magnet and spin-polarized electrons in a simple model simulation not only induce the STT effect but also generate quantum fluctuations due to the quantum disentanglement process.\cite{MonCar} In this paper, we will show that the quantum macrospin scattering model we exploited before is exactly solvable, and will investigate the magnetization dynamics from the full quantum picture by applying the standard theoretical techniques for open quantum systems to this model.          

\section{Quantum Macrospin Model}
In parallel with the original study of STT in the semiclassical picture\cite{torq1,torq2}, we consider the motion of a single-domain magnet driven by the spin-polarized current. However, the magnet here is not described by the classical magnetization vector $\mathbf{M}$, but is represented by the quantum operators of the total spin angular momentum $\widehat{\mathbf{J}}$. The spin-polarized electrons are injected along the $x$-direction in sequence independent of one another, and interact with the magnet located at $x=0$ through the exchange interaction. The model Hamiltonian for each electron interacting with the magnet is\cite{MonCar}
\begin{equation}
H=-\frac{1}{2}\partial_{x}^{2}+\delta(x)\left(\lambda_{0}\;
\widehat{\mathsf{J}}_{0}+\lambda\; \widehat{\mathbf{s}}\cdot\widehat{\mathbf{J}}\right). \label{Hamil2}
\end{equation}
where the first term is the kinetic term of a single current electron, and the second term is the interaction between the electrons and the magnet; $\widehat{\mathsf{J}}_{0}$ and $\widehat{\mathbf{J}}$ are the unit and total spin operators of the magnet, and $\widehat{\mathbf{s}}$ is the electron spin operator; the parameters $\lambda_{0}$ and $\lambda$ are the spin-independent and spin-dependent interaction strength respectively, which are used in the semiclassical model if the operators $\widehat{\mathsf{J}}_{0}$ and $\widehat{\mathbf{J}}$ are replaced by their mean field approximation according to the correspondence principle. Note that the Hamiltonian is that of a free magnet without the external magnetic field and anisotropic crystal field. This treatment will simplify the calculations below without invalidating of the general conclusions. 

The STT effect originates from the elementary entangle-disentangle processes of the scattering states between the magnet and the spin-polarized electrons.\cite{MonCar} These scattering states are deduced from the scattering matrix $\mathcal{S}$ of the Hamiltonian (\ref{Hamil2}). Unlike the scattering matrix in the semiclassical picture, which is defined only in the Hilbert space of the electron, the scattering matrix $\mathcal{S}$ in this full quantum picture is defined on the larger Hilbert space including both the magnet and the electron (see Appendix A), which gives the STT directly and more informations compared with the semiclassical case.

\section{Quantum Dynamics Equations for Magnet}
\subsection{Quantum Master Equation}
Consider the scattering of the magnet spins by an injected electron as uncorrelated. The incoming composite state of the magnet and the current electron is assumed to be a product of their respective  density matrices $\rho_{\text{in}}^{J}$ and $\rho_{\text{in}}^{e}$. After  scattering, the outgoing states of the whole system, $\rho_{\text{out}}=\mathcal{S}\rho_{\text{in}}^{J}\otimes \rho_{\text{in}}^{e}\mathcal{S}^{\dag}$, as a result of the unitary scattering matrix $\mathcal{S}$, has a degree of entanglement. Next, the surrounding environment decoheres the entangled state into a joint probability distribution of   the possible magnet states and the corresponding electron states. Properties of the resultant magnet state or the electron state are characterized by their respective reduced density  $\rho_{\text{out}}^{J}$ or electron $\rho_{\text{out}}^{e}$ from tracing over the degrees of freedom of the other component in $\rho_{\text{out}}$. The \textit{correlated} quantum dynamics of the magnet and the electron injected in sequence in the spin-polarized current drives the time evolution of the magnet state $\rho^{J}$ and the magnetization-dependent electron transport properties in the electron density matrix $\rho^{e}$. While the mean magnetization dynamics is within reach of the semiclassical picture, the magnetization fluctuation is given only by the full quantum treatment followed here without additional stochastic assumptions.

For a theory of dynamics of the ferromagnet as an open system, we treat the current electrons as the equivalent of the environment. The Kraus operator\cite{Kraus} of the magnet is defined in terms of the scattering matrix $\mathcal{S}$ as the evolution operator of each encounter with a current electron,
\begin{eqnarray}
\mathcal{K}_{k,s;k',s'}\equiv\langle k,s|\mathcal{S}|k',s'\rangle,  \label{Smatrix}
\end{eqnarray}
with a specific basis set $\{|k,s\rangle \}$ of an incoming electron state of wave vector $k$ and spin up or down state $s=\pm$. We have adopted the simple model (\ref{Hamil2}) for the dynamics of the rigid macro-spin states $\{ |J,m\rangle\}$ with the total spin number $J$ and the $z$ component quantum number $m$ and leave the effects of spin waves for future study. Then, the current electron kinetic energy is conserved and the Kraus operators are non-zero only if $k$ and $k'$ are on the same energy shell, given by 
\begin{eqnarray}
\mathcal{K}_{k,s;\pm k,s}=(\xi\pm\frac{1}{2})\widehat{\mathsf{J}}_{0}+s\zeta\widehat{\mathsf{J}}_{z},\quad
\mathcal{K}_{k,-s;\pm k,s}=\zeta\widehat{\mathsf{J}}_{s},  \label{KrausOp}
\end{eqnarray}
where the coefficients $\xi$ and $\zeta$ are functions of the basic parameters $\lambda_{0},\lambda$ and $k$, $J$. The first operator  $\mathcal{K}_{k,s;\pm k,s}$ with the same spin index $s$ comes from  scattering without spin transfer, and the second operator $\mathcal{K}_{k,-s;\pm k,s}$ with a change in $s$ represents spin transfer. These Kraus operators are functions of the macro-spin $\widehat{\mathsf{J}}_{0}$ and $\widehat{\mathbf{J}}$ (see Appendix B). In the semiclassical picture, these Kraus operators will reduce to scalars representing effective fields for the magnetization dynamics. 

In a scattering event, let the initial state of the current electron be given by the density matrix 
$\rho_{\text{in}}^{e}=\sum_{s,s'}f_{s,s'}(k)|k,s\rangle\langle k,s'|$. This simple form may be extended to account for a wave vector distribution or quantum coherence between different wave vectors, but will not be exploited here to keep the exposition simple.  With the above Kraus operators, the quantum map of the magnet state from $\rho_{\text{in}}^{J}$ to $\rho_{\text{out}}^{J}$ in the scattering  is 
\begin{eqnarray}
\rho_{\text{out}}^{J}=\sum_{\pm,s,s',s''}f_{s,s'}(k)\mathcal{K}_{\pm k,s'';k,s}\rho_{\text{in}}^{J}(\mathcal{K}_{\pm k,s'';k,s'})^{\dag}. \label{DynMap}
\end{eqnarray} 
If the spin-polarized current is considered as a sequence of electrons injected at equal time interval $\tau$ (a measure of the inverse current), the continuous evolution of the magnetic density matrix driven by Eq.~(\ref{DynMap}), with a coarse graining of a time scale much larger than  $\tau$,  yields the quantum dynamics of the magnet governed by the master equation,
\begin{equation}
\frac{\partial}{\partial t}\rho^{J}(t)=\frac{1}{\tau}[\mathcal{T}_{0}(t)+\mathbf{S}(t)\cdot \boldsymbol{\mathcal{T}}(t)],  \label{Master}
\end{equation}
where $\mathbf{S}=\text{Tr}[\boldsymbol{\sigma}\rho_{\text{in}}^{e}]$ is the Bloch vector of the spin-polarized current electrons,  $\boldsymbol{\sigma}$ being the Pauli matrices. The operators $\mathcal{T}_{0}$ and $\boldsymbol{\mathcal{T}}$ are polynomial functions of $\widehat{\mathsf{J}}_{0}$ and $\widehat{\mathbf{J}}$ (see Appendix C). $\mathcal{T}_{0}$ corresponds to the unpolarized part of the current which causes fluctuations of the magnet motion without a net spin transfer effect. On the other hand, $\boldsymbol{\mathcal{T}}$ is induced by the electron spin polarization, giving rise to both STT and the magnetization fluctuations.    

Note that the master equation (\ref{Master}) is an exact solution from the model (\ref{Hamil2}) for arbitrary $J$. Thus, Eq.~(\ref{Master}) can  be  applied to molecular magnets of small $J$.  This is in contrast with the simulation of the  quantum stochastic dynamics of the magnet in a previous study\cite{MonCar} which used the same scattering matrix but required the large $J$ condition  to keep the approximation of the magnetic quantum state as a spin coherent state after scattering.   

\subsection{Fokker-Planck Equation}
To facilitate computation in the large $J$ case and, more importantly, to study the quantum-classical crossover of the magnetization dynamics, we put the  master equation (\ref{Master}) in the spin coherent state  $\mathcal{P}$-representation\cite{OpenQuan1,arecchi72} analogous to the boson case. The chosen basis set is the overcomplete and non-orthogonal states $\{|J,\Omega\rangle\}$, where $|J,\Omega\rangle$ is the spin coherent state of total spin $J$ in the direction of  $\Omega=(\Theta,\Phi)$. The density matrix $\rho^{J}$ in this representation is  $\rho^{J}(t)\equiv\int d\Omega\mathcal{P}_{J}(\Omega,t)|J,\Omega\rangle\langle J,\Omega|$, and the spin operators $\widehat{\mathbf{J}}$ take the form of the differential operators.\cite{Dalgebra} Substituting these expressions of $\rho^{J}$ and $\widehat{\mathbf{J}}$ into Eq.~(\ref{Master}) with some algebraic manipulations, we obtain the Fokker-Planck equation for $\mathcal{P}_{J}$ (see Appendix D),
\begin{eqnarray}
\frac{\partial}{\partial t}\mathcal{P}_{J}(\widehat{\mathbf{m}},t)
=-\nabla\cdot(\mathbf{T}\mathcal{P}_{J})+\nabla^{2}(\mathcal{D}\mathcal{P}_{J}), \label{FPE2}
\end{eqnarray}  
where the unit vector $\widehat{\mathbf{m}}$ points in the direction of the macrospin $\Omega=(\Theta,\Phi)$, the drift vector $\mathbf{T}=\mathcal{A}(\widehat{\mathbf{m}}\times\mathbf{S})\times\widehat{\mathbf{m}}
+\mathcal{B}\widehat{\mathbf{m}}\times\mathbf{S}$ contains the two well-known terms of 
STT,\cite{revtorq1,revtorq2} the diffusion coefficient $\mathcal{D}=\mathcal{A}(1-\widehat{\mathbf{m}}\cdot\mathbf{S})/(2J+1)$ originates from the quantum fluctuation generated by the scattering.\cite{MonCar} The parameters $\mathcal{A}$ and $\mathcal{B}$  are functions of the parameters $\xi$ and $\zeta$ in the Kraus operators (\ref{KrausOp}), namely, $\mathcal{A}=(2J+1)|\zeta|^{2}/\tau,\quad\mathcal{B}=2\Im[\xi^{\ast}\zeta]/\tau$ ($\Im$ denoting the imaginary part of) which can be determined from the basic parameters of the Hamiltonian (\ref{Hamil2}).

The quasi-probability distribution function $\mathcal{P}_{J}$ in Eq.~(\ref{FPE2}) is different from the one considered in the classical theory. Its value could be negative in some situations because $\mathcal{P}_{J}$ describes the quantum state of the magnet as faithfully as the density matrix $\rho^{J}$. As shown in Eq.~(\ref{FPE2}), the STT terms naturally arise from the open quantum dynamics of the magnet in the presence of the continuous scatterings by the spin-polarized electrons. Unlike the semiclassical picture, where the STT terms are indirectly obtained from the current electron spin polarization after potential scattering, in the quantum case, the STT terms follows directly from the full quantum scattering. Furthermore, the full quantum treatment also gives the quantum fluctuations accompanying the spin transfer processes, which does not exist in the semiclassical treatment.

The diffusion coefficient $\mathcal{D}$ expression shows dependence on the relative angle between the magnet and the electron spin, with its maximal (minimal) value when $\widehat{\mathbf{m}}$ and $\mathbf{S}$ are anti-parallel (parallel). This coincides with our previous simulation\cite{MonCar} that the quantum magnetization fluctuation is first enhanced and then suppressed during the STT-driven magnet switching, which is  observed in a recent experiment.\cite{CurrFluc} For the special case where the injected electrons are fully unpolarized ($\mathbf{S}=\mathbf{0}$), $\mathcal{D}$ is still a non-zero constant which again suggests that the unpolarized current can cause quantum magnetization fluctuations without net spin transfer. The steady solution of Eq.~(\ref{FPE2}) is a constant, which means a uniform distribution function in the spin coherent state space and the magnetization will vanish on average. By contrast, the semiclassical theory   predicts only the zero spin torque but no diffusion dynamics for the magnet.
This STT-induced magnetization fluctuation becomes dominant over the thermal magnetization fluctuation at low temperatures. The crossover temperature is estimated by comparing the diffusion coefficient $\mathcal{D}$ in Eq.~(\ref{FPE2}) with the one for thermal magnetization fluctuation,\cite{Brown,Therm} 
\begin{equation}
\frac{\alpha_{g}\gamma_{g}k_{B}T}{|\mathbf{M}|}\sim\frac{\mathcal{A}}{2J+1}(1-\widehat{\mathbf{m}}\cdot\mathbf{S}),
\end{equation}
where $\mathbf{M}$ is the magnetic moment of the magnet, $\alpha_{g}$  the Gilbert damping coefficient, $\gamma_{g}$  the gyromagnetic ratio, $k_{B}$  the Boltzmann constant, and $T$  the temperature. Since $|\mathbf{M}|=\gamma_{g}J\hbar$ and $|\zeta|^{2}\sim\mathcal{O}(1/J^{2})$ for $J\gg1$, we obtain $\alpha_{g}\kappa_{B}T \sim \hbar/J\tau$, in agreement with the quantum noise estimate.\cite{MonCar}

The $\mathcal{P}_{J}$ distribution as the solution of the Fokker-Planck equation (\ref{FPE2}) gives the exact quantum dynamics of the magnet based on the quantum macrospin model (\ref{Hamil2}). The expectation values of any observable physics qualities, such as the magnetization and its fluctuations, can be calculated from the $\mathcal{P}_{J}$ distribution. An example is demonstrated in subsection D. In the semiclassical picture, the STT is defined on the mean field level of the magnetization, and quantum correlation between the magnetization states does not exist. The quantum theory includes the quantum correlation and is applicable to any possible exotic quantum states of the magnet in the mesoscopic or microscopic regime. 

\subsection{WKB Approximation for Large $J$}
In the large  $J$ regime, we demonstrate that the solution of Eq.~(\ref{FPE2}) leads to classical behavior.  The expressions for $\mathcal{A}$ and $\mathcal{B}$ suggest that $\mathbf{T}\sim\mathcal{O}(1/J)$ and $\mathcal{D}\sim\mathcal{O}(1/J^{2})$, thus the diffusion term is smaller than the drift term in Eq.~(\ref{FPE2}) by the order of magnitude $\mathcal{O}(1/J)$. Then, the WKB approximation is applied for large $J$.\cite{WKB} Substituting $\mathcal{P}_{J}(\widehat{\mathbf{m}},t)=e^{-JW(\widehat{\mathbf{m}},t)}$ into Eq.~(\ref{FPE2}) and keeping the terms to the leading order of $1/J$, leads to the Hamilton-Jacobi equation for $W(\widehat{\mathbf{m}},t)$,
\begin{equation}
\frac{\partial W}{\partial t}+\mathbf{T}\cdot\nabla W+J\mathcal{D}(\nabla W)^{2}=0.\label{HamJoc}
\end{equation}
Thus, the STT-driven magnet obeys the canonical dynamics in the classical limit, and the function $W$ plays the role of action. For the constant spin-polarized current case, the corresponding Hamiltonian canonical equations in the spherical polar coordinates are
\begin{eqnarray}
\frac{d\Theta}{dt}&=&\frac{\partial\mathcal{H}}{\partial P_{\Theta}},\quad
\frac{dP_{\Theta}}{dt}=-\frac{\partial\mathcal{H}}{\partial\Theta},\nonumber\\
\frac{d\Phi}{dt}&=&\frac{\partial\mathcal{H}}{\partial P_{\Phi}},\quad
\frac{dP_{\Phi}}{dt}=-\frac{\partial\mathcal{H}}{\partial\Phi}.\label{HamCan}
\end{eqnarray}  
Here, the Hamiltonian function is explicitly written as
\begin{eqnarray}
\mathcal{H}=\mathsf{T}_{\Theta}P_{\Theta}+\frac{\mathsf{T}_{\Phi}}{\sin\Theta}P_{\Phi}
+J\mathcal{D}P_{\Theta}^{2}+\frac{J\mathcal{D}}{\sin^{2}\Theta}P_{\Phi}^{2}. \label{HamFun}
\end{eqnarray}
with the generalized momentum components, $P_{\Theta}=\partial_{\Theta}W$ and $P_{\Phi}=\partial_{\Phi}W$, and the STT components $\mathsf{T}_{\Theta}$ and $\mathsf{T}_{\Phi}$. The equations for $\Theta$ and $\Phi$ in (\ref{HamCan}) show that two more terms, which originate from the diffusion term in Eq.~(\ref{FPE2}), contribute to the magnetic dynamics in addition to the STT terms even in the classical limit. Eq.~(\ref{HamCan}) and (\ref{HamFun}) give a more complete description of the classical magnetization dynamics than the semiclassical STT theory. 

\subsection{Numerical Example}
\begin{figure}[tbp]
\subfigure{
    \begin{minipage}{0.225\textwidth}
      \flushleft
      \includegraphics[scale=0.15]{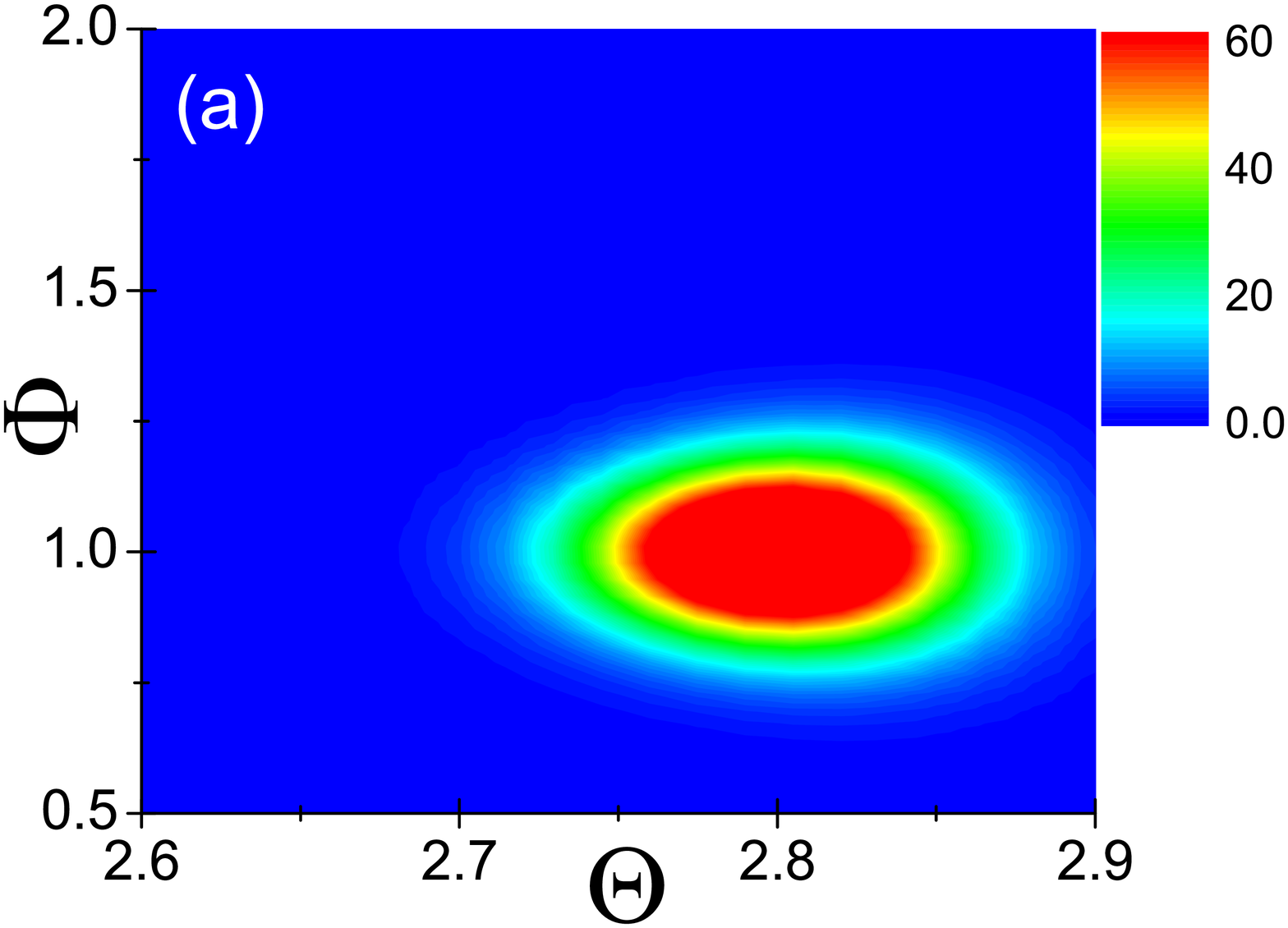}
    \end{minipage}}
\subfigure{
    \begin{minipage}{0.225\textwidth}
      \flushleft
      \includegraphics[scale=0.15]{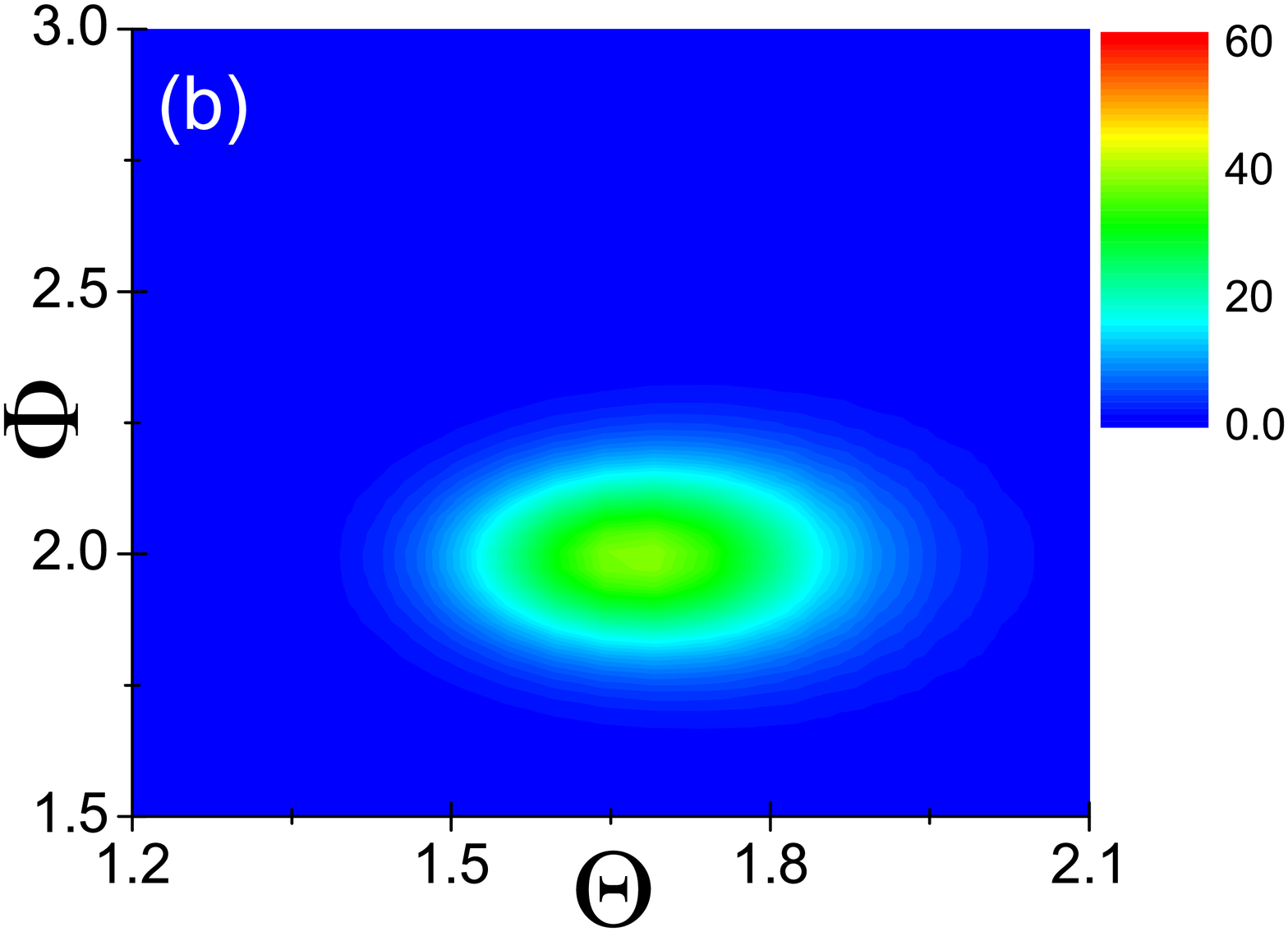}
    \end{minipage}}
\subfigure{
    \begin{minipage}{0.225\textwidth}
      \flushleft
      \includegraphics[scale=0.15]{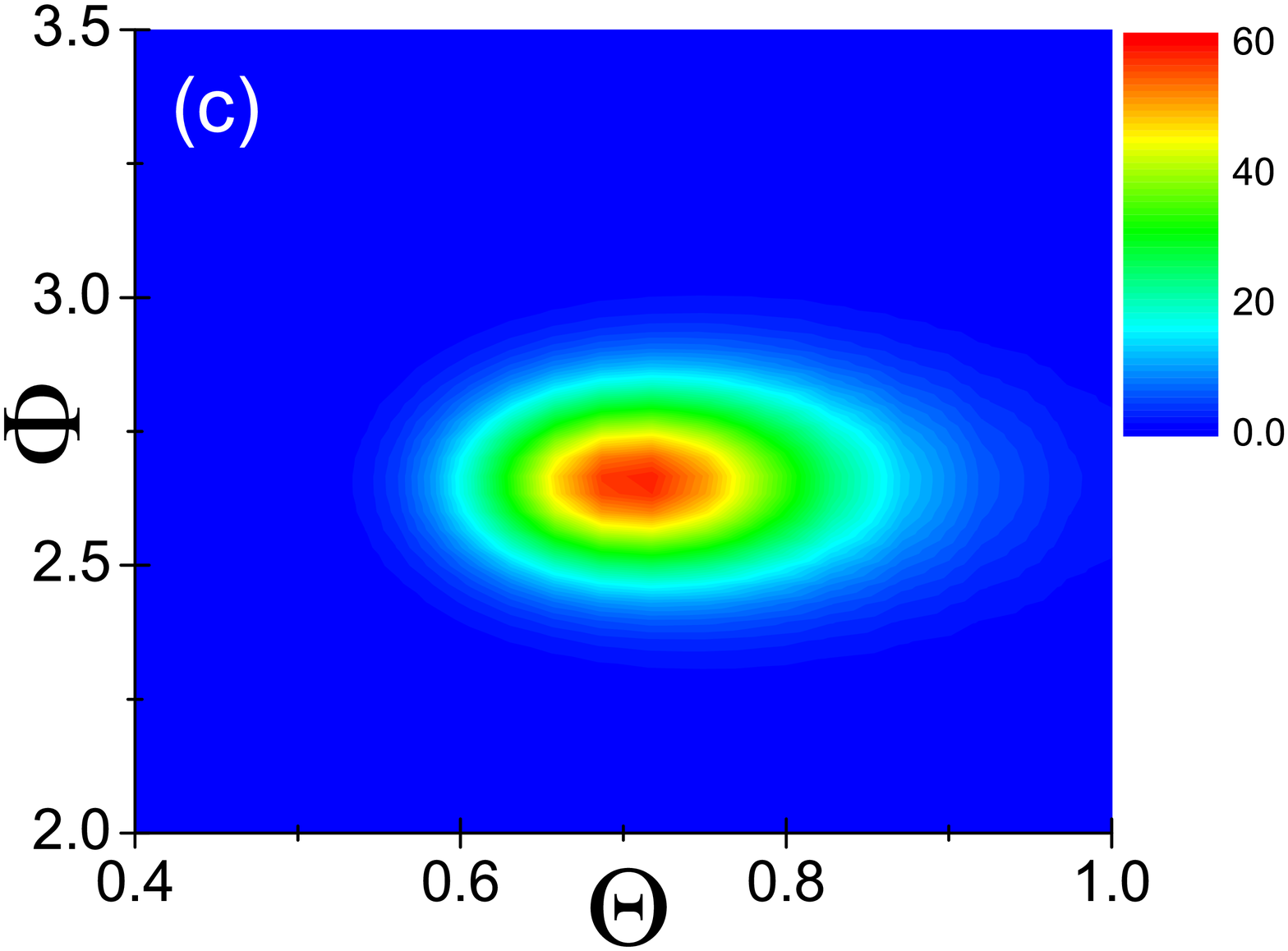}
    \end{minipage}}
\subfigure{
    \begin{minipage}{0.225\textwidth}
      \flushleft
      \includegraphics[scale=0.15]{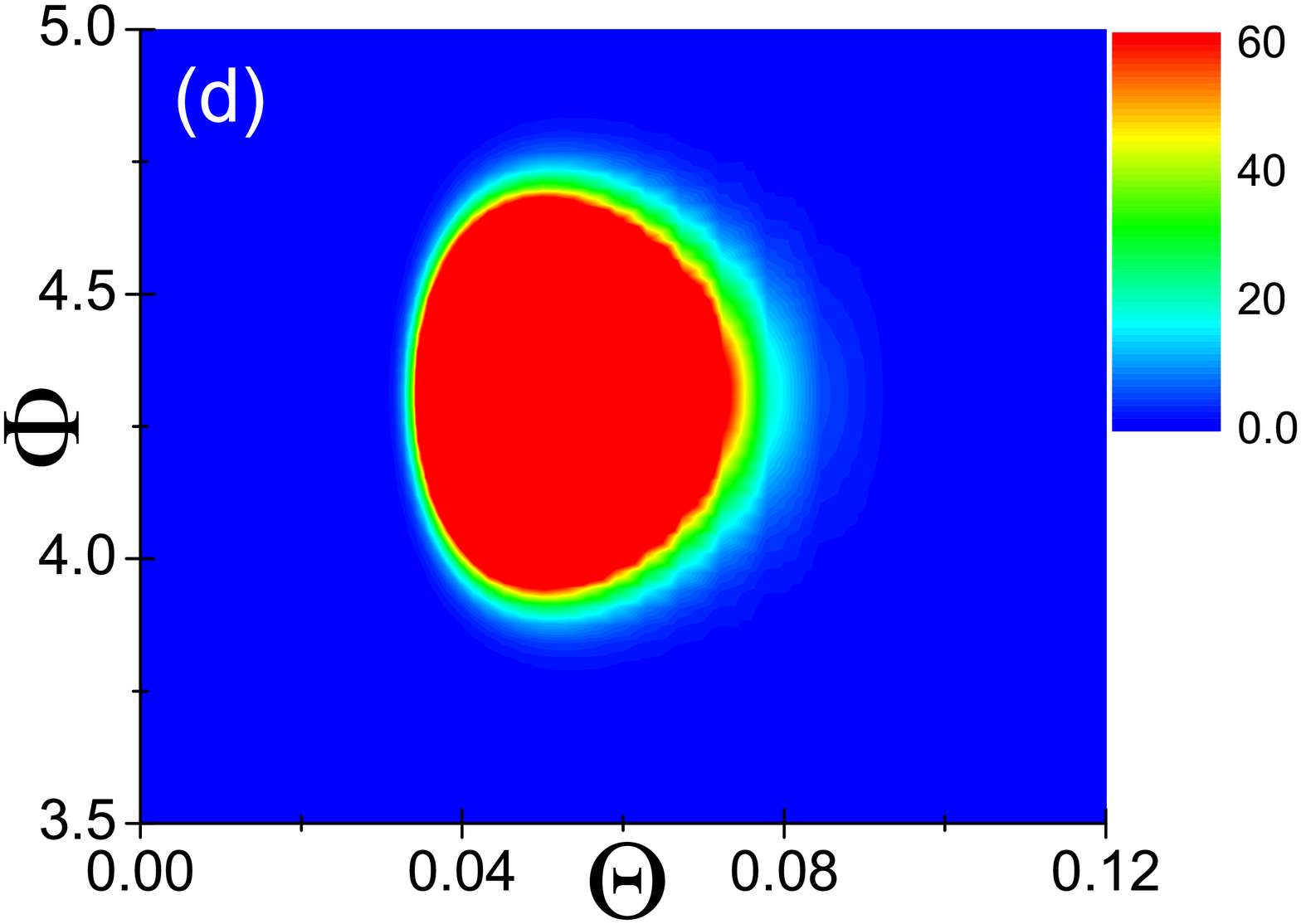}
    \end{minipage}}
\caption{(color online). Distribution function $\mathcal{P}_{J}$ for the nanomagnet at four different time $t$. (a) $t=0$; (b) $t=0.3$~$t_{N}$; (c) $t=0.5$~$t_{N}$;
 (d) $t=t_{N}$. The simulation parameters are set as: $J=10^{4}$, $\Delta_{+}=1.3$~V, $\Delta_{-}=0.1$~V, $d=3$~nm, $N=1.5\times10^{5}$, $k=13.6$~nm$^{-1}$, $\mathbf{S}=(0,0,1)$. A 200$\times$200 lattice in $\Theta$-$\Phi$ plane is used in the simulations.}
\label{Pdist}
\end{figure}

Here we demonstrate how to apply the approach developed above to the STT-driven quantum dynamics of a magnet. We consider a magnet with $J=10^{4}$, and the initial distribution function obeys the Boltzmann distribution, i.e. $\mathcal{P}_{J}(\widehat{\mathbf{m}},0)=Ce^{-E/k_{B}T}$, where the energy of the magnet in a magnetic field $\mathbf{B}$ is $E=-\mathbf{M}\cdot\mathbf{B}$, $C$ is the normalization factor, and the magnetic moment is $\mathbf{M}=\gamma_{g}\hbar J\widehat{\mathbf{m}}$. We set the temperature as $T=1$~K, and the magnitude of the magnetic field $B=0.05$~T. The direction of $\mathbf{B}$ is chosen that maximum value of $\mathcal{P}_{J}$ locates at the angle $\Omega_{0}=(2.8,1.0)$. The initial distribution is show in the Fig.~\ref{Pdist}(a). Then we apply a spin current pulse, which includes $1.5\times 10^{5}$ electrons. The Bloch vector of the electron spin is $\mathbf{S}=(0,0,1)$, and the wavevector is $k=13.6$~nm$^{-1}$. To calculate the scattering matrix, the parameters $\lambda_{0}$ and $\lambda$ in Eq.~(\ref{Hamil2}) are estimated for a magnet with effective potentials $\Delta_{+}=1.3$~V, $\Delta_{-}=0.1$~V and thickness $d=3$~nm. 

In the simulations, we have used the method of characteristics\cite{Method} to solve the Hamilton-Jacobi equation (\ref{HamJoc}), which gives the time-evolution of $W(\widehat{\mathbf{m}},t)$ and then the distribution function $\mathcal{P}_{J}(\widehat{\mathbf{m}},t)=e^{-JW(\widehat{\mathbf{m}},t)}$. Direct solution of the Fokker-Planck equation (\ref{FPE2}) or the master equation (\ref{Master}) is also practical but not explored here. The time is measured in $t_{N}=N\tau$. The distribution functions  $\mathcal{P}_{J}$ at $t=0.3t_{N},0.5t_{N},t_{N}$ are shown in Fig.~\ref{Pdist} (b)(c)(d) respectively. In order to keep the normalization of $\mathcal{P}_{J}$, it is renormalized after every 500 scatterings. Note the different scales for $\Theta$ in Fig.~\ref{Pdist}. We found that $\mathcal{P}_{J}$ is first expanded and then compressed, and its center moves from (2.8,1.0) to (0.05,4.2), which show the effect of spin current on the distribution function.   

\begin{figure}[tbp]
\subfigure{
    \begin{minipage}{0.225\textwidth}
      \flushleft
      \includegraphics[scale=0.16]{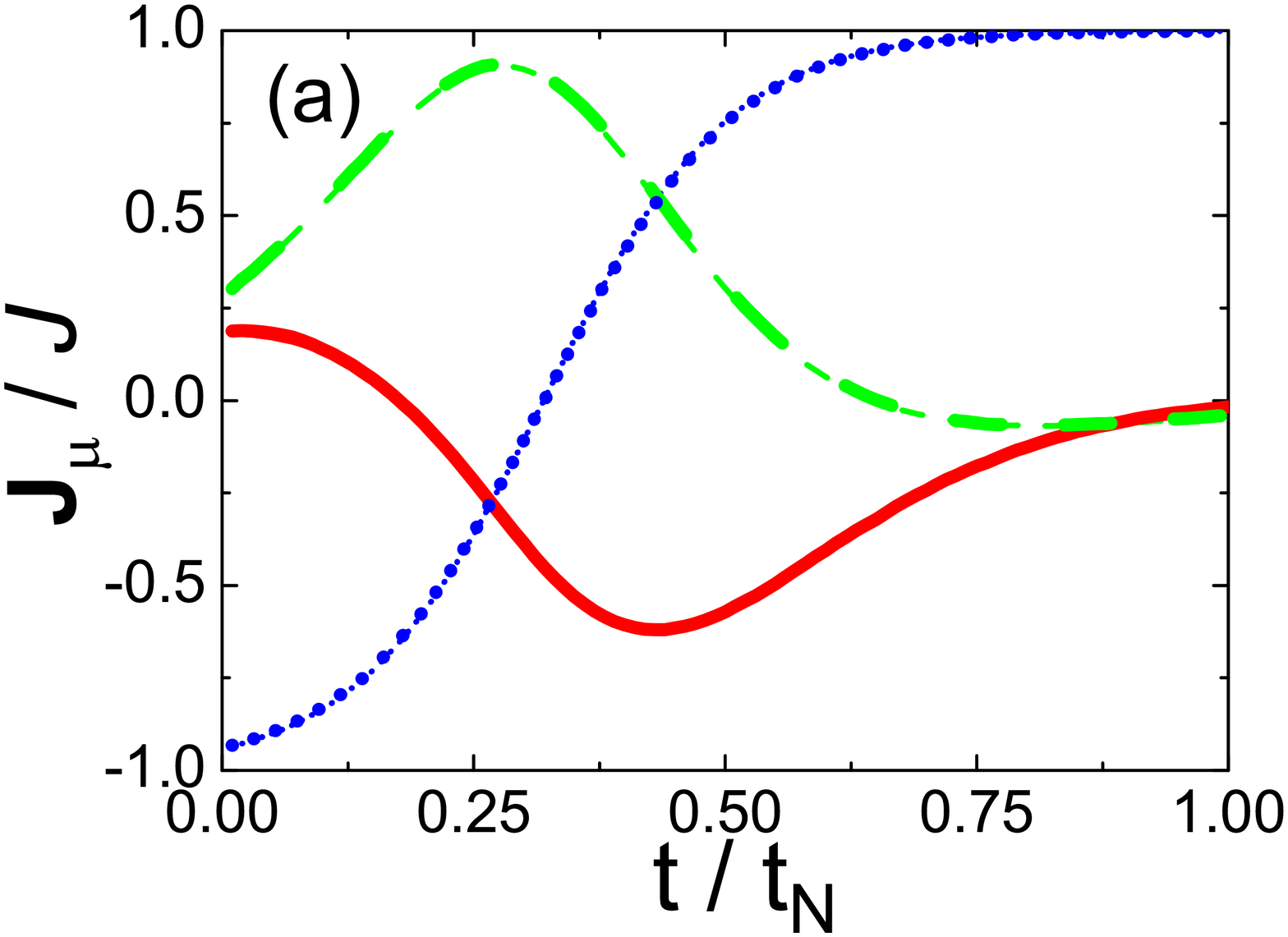}
    \end{minipage}}
\subfigure{
    \begin{minipage}{0.225\textwidth}
      \flushleft
      \includegraphics[scale=0.16]{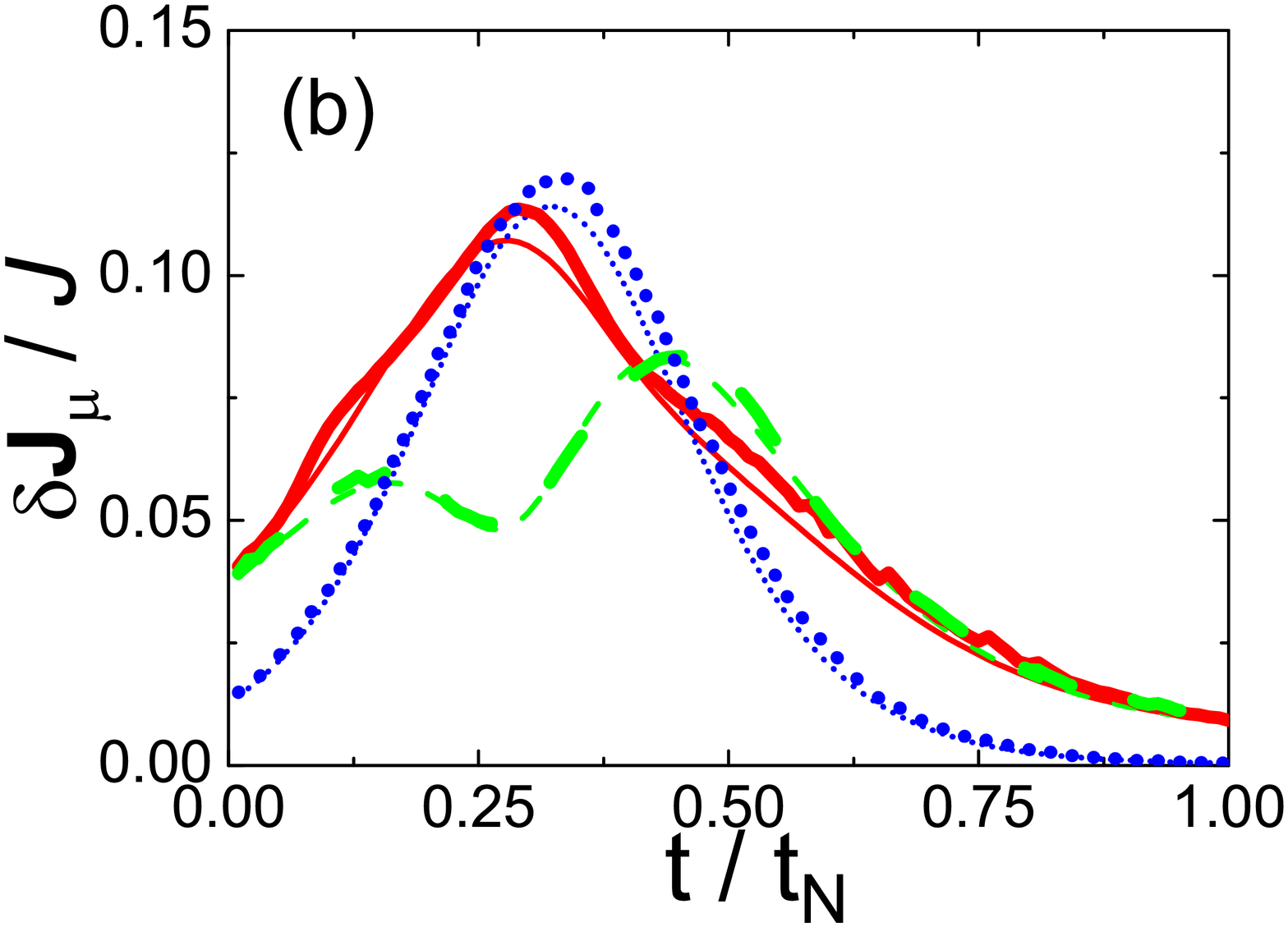}
    \end{minipage}}
\caption{(color online). Time-evolution of $\mathsf{J}_{\mu=x,y,z}$ and its fluctuation
$\delta\mathsf{J}_{\mu=x,y,z}$~($x$: solid line; $y$: dash line; $z$: dotted line.) caused by the spin transfer torque obtained from the probability distribution function~(thick line) and the quantum trajectory method~(thin line). }
\label{MdM}
\end{figure}

The mean value of the macrospin $\widehat{\mathbf{J}}$ and its fluctuations are calculated from $\mathcal{P}_{J}$ as
\begin{eqnarray*}
\mathsf{J}_{\mu=x,y,z}(t)&=&\int d\Omega\mathcal{P}(\Omega,t)\langle J,\Omega |\widehat{\mathsf{J}}_{\mu}|J,\Omega\rangle,\\
\mathsf{\delta J}_{\mu=x,y,z}^{2}(t)&=&\int d\Omega\mathcal{P}(\Omega,t)\langle J,\Omega|\delta\widehat{\mathsf{J}}_{\mu}^{2}|J,\Omega\rangle.
\end{eqnarray*}
The results are shown in Fig.~\ref{MdM}, where the mean magnetization is switched by the STT, and the magnetization fluctuations are first enhanced and finally suppressed. Comparison with the results obtained from the quantum trajectory method in Ref.~\onlinecite{MonCar} gives reasonable agreement. The approach developed here together with the quantum trajectory method\cite{MonCar} fits in the toolbox for micromagnetics simulations with the added value of accounting for relevant quantum effects.  
 
\section{Electron Density Matrix After Scattering}
Finally, we briefly discuss the electron states after the scattering, which contain the informations about the electric current and current noise, etc. After tracing over the degrees of freedom for the magnet in the total density matrix $\rho_{out}$, the reduced density matrix of the electron is $\rho_{out}^{e}=\mathbf{Tr}_{J}[\mathcal{S}\rho_{in}^{e}\otimes\rho_{in}^{J}\mathcal{S}^{\dag}]$. This expression is a generalization of the electron potential scattering in the semiclassical picture\cite{SemiScat}. Here, the transformation of the electron state is no longer unitary due to the recoil of the magnet. For instance, we assume that one electron with wavevector $k>0$ and spin-polarized vector $\mathbf{S}=(0,0,1)$ is injected, i.e., $\rho_{in}^{e}=|k,+\rangle\langle k,+|$, and take the $\mathcal{P}$-representation for the quantum states of the magnet. With the scattering matrix $\mathcal{S}$, we obtain
\begin{eqnarray}
\rho_{out}^{e}&=&\sum_{k',s';k'',s''}|k',s'\rangle\langle k'',s''|\nonumber\\&\times&
\int d\Omega\mathcal{P}(\Omega)\langle\Omega|(\mathcal{K}_{k'',s'';k,+})^{\dag}\mathcal{K}_{k',s';k,+}|\Omega\rangle.\label{erho}
\end{eqnarray}
For the model (\ref{Hamil2}), the terms in Eq.~(\ref{erho}) are non-zero only if $k'=\pm k$ and $k''=\pm k$. The electron scattering state is correlated with the quantum state of the magnet. The scattering formalism in the semiclassical picture will be reproduced if the Kraus operators are replaced by the corresponding scattering matrix elements there. As the magnets are miniaturized further\cite{Tbit} and the quantum description becomes necessary, the semiclassical scattering formalism for the electron transport will break down, and Eq.~(\ref{erho}) and its generalizations should be exploited as the new starting point.

\section{Conclusion}
In conclusion, the STT-driven magnetization dynamics has been investigated by treating the magnet as an open quantum system in the exactly solvable quantum macrospin model. A set of dynamical equations is established and the quantum-classical connection is made. The full quantum picture here provides a unified and complete description of the magnetization dynamics and electron transport, and further explorations of the quantum physics in spintronics along this line is expected. 

\begin{acknowledgements}
We acknowledge the support of this work by the U. S. Army Research Office under contract number ARO-MURI W911NF-08-2-0032 and NSF ECCS-1202583.
\end{acknowledgements}

\appendix
\section{Scattering Matrix}
First, we calculate the scattering matrix $\mathcal{S}$ of the model Hamiltonian
\begin{equation}
H=-\frac{1}{2}\partial_{x}^{2}+\delta(x)\left(\lambda_{0}\;
\widehat{\mathsf{J}}_{0}+\lambda\widehat{\mathbf{s}}\cdot\widehat{\mathbf{J}}\right). \label{SHamil2}
\end{equation}
Considering that one electron in state $|\psi_{in}^{e}\rangle$ is injected along the $x$-direction and the initial quantum state of the magnet is $|\psi_{in}^{J}\rangle$, then the incoming state $|\Psi_{in}\rangle$ of the whole system before scattering is the product state of $|\psi_{in}^{e}\rangle$ and $|\psi_{in}^{J}\rangle$, i.e., $|\Psi_{in}\rangle=|\psi_{in}^{e}\rangle\otimes|\psi_{in}^{J}\rangle$. After scattering, the outgoing state $|\Psi_{out}\rangle$ will be $|\Psi_{out}\rangle=\mathcal{S}|\Psi_{in}\rangle$. The scattering matrix $\mathcal{S}$ are determined by the boundary conditions at $x=0$, 
\begin{eqnarray}
(\Psi_{in}+\Psi_{out})|_{x=0^{-}} &=&(\Psi_{in}+\Psi_{out})|_{x=0^{+}}, \nonumber\\
\int_{0^{-}}^{0^{+}}H(\Psi_{in}+\Psi_{out})dx &=&\varepsilon
\int_{0^{-}}^{0^{+}}(\Psi_{in}+\Psi_{out})dx,\label{SScatProb}
\end{eqnarray}
where $\varepsilon$ is the total energy of the whole system.

The scattering problem (\ref{SScatProb}) is simplified by utilizing the symmetries of the model Hamiltonian (\ref{SHamil2}). First, the kinetic energy of the electron is conserved during the scattering process. Thus the scattering matrix elements of $\mathcal{S}$ is non-zero only if the absolute values of the incoming and outgoing wavevectors of the electron are the same. Second, the operators $(\widehat{\mathbf{s}}+\widehat{\mathbf{J}})^{2}$ and $\widehat{\mathsf{s}}_{z}+\widehat{\mathsf{J}}_{z}$ are commutative with the Hamiltonian (\ref{SHamil2}), and their eigenstates $|\mathcal{J},\mu\rangle$ are given as
\begin{eqnarray*}
(\widehat{\mathbf{s}}+\widehat{\mathbf{J}})^{2}|\mathcal{J},\mu\rangle&=&
\mathcal{J}(\mathcal{J}+1)|\mathcal{J},\mu\rangle,\\ (\widehat{\mathsf{s}}_{z}+\widehat{\mathsf{J}}_{z})|\mathcal{J},\mu\rangle&=&\mu|\mathcal{J},\mu\rangle,
\end{eqnarray*} 
with $\mathcal{J}=J\pm\frac{1}{2}$ and $\mu=-\mathcal{J},...,\mathcal{J}$. Choosing the basis set $\{|k;\mathcal{J},\mu\rangle\}$, the scattering problem (\ref{SScatProb}) reduced to a set of $\delta$-potential scattering equations, and gives the scattering matrix $\mathcal{S}$ in this representation\cite{MonCar}. Then after a representation transformation with the help of Clebsch-Gorden coefficients, we obtain the scattering matrix $\mathcal{S}$ in the basis set $\{|k,s;J,m\rangle\}$, which takes a block form
\begin{eqnarray}
\mathcal{S}=\left(
\begin{array}{ccc}
\ddots & 0 & 0  \\
0 & \mathcal{S}_{k,\mu}  & 0 \\
0 & 0 & \ddots
\end{array}\right),\label{SScatMax}
\end{eqnarray}
and the form of each block $\mathcal{S}_{k,\mu }$ is  
\begin{equation}
\mathcal{S}_{k,\mu}=\left(
\begin{array}{cccc}
t_{k,\mu }^{++} & r_{k,\mu }^{++} & t_{k,\mu }^{+-}& r_{k,\mu }^{+-} \\
r_{k,\mu }^{++} & t_{k,\mu }^{++} & r_{k,\mu }^{+-}& t_{k,\mu }^{+-} \\
t_{k,\mu }^{-+} & r_{k,\mu }^{-+} & t_{k,\mu }^{--}& r_{k,\mu }^{--} \\
r_{k,\mu }^{-+} & t_{k,\mu }^{-+} & r_{k,\mu }^{--}& t_{k,\mu }^{--}
\end{array}\right).  \label{SScatBlock}
\end{equation}
Here, the element $t_{k,\mu}^{ss'}$ ($r_{k,\mu}^{ss'}$) is the transmission (reflection) probability amplitude from $|k,s';J,\mu-\frac{1}{2}s'\rangle$ to $|k,s;J,\mu-\frac{1}{2}s\rangle$ ($|-k,s;J,\mu-\frac{1}{2}s\rangle$). The spin transfer is related to those elements with $s\neq s'$. The explicit expressions for the matrix elements are
\begin{small}
\begin{eqnarray*}
t_{k,\mu }^{++} &=&\cos \eta _{J,+}e^{-i\eta _{J,+}}\cos ^{2}\alpha
_{J,\mu }+\cos \eta _{J,-}e^{-i\eta _{J,-}}\sin ^{2}\alpha _{J,\mu }, \\
r_{k,\mu }^{++} &=&-i(\sin \eta _{J,+}e^{-i\eta _{J,+}}\cos
^{2}\alpha _{J,\mu }+\sin \eta _{J,-}e^{-i\eta _{J,-}}\sin ^{2}\alpha_{J,\mu }), \\
t_{k,\mu }^{--} &=&\cos \eta _{J,+}e^{-i\eta _{J,+}}\sin ^{2}\alpha
_{J,\mu }+\cos \eta _{J,-}e^{-i\eta _{J,-}}\cos ^{2}\alpha _{J,\mu }, \\
r_{k,\mu }^{--} &=&-i(\sin \eta _{J,+}e^{-i\eta _{J,+}}\sin
^{2}\alpha _{J,\mu }+\sin \eta _{J,-}e^{-i\eta _{J,-}}\cos ^{2}\alpha_{J,\mu }), \\
t_{k,\mu }^{-+} &=&(\cos \eta _{J,+}e^{-i\eta _{J,+}}-\cos \eta
_{J,-}e^{-i\eta _{J,-}})\sin \alpha _{J,\mu }\cos \alpha _{J,\mu }, \\
r_{k,\mu }^{-+} &=&-i(\sin \eta _{J,+}e^{-i\eta _{J,+}}-\sin \eta
_{J,-}e^{-i\eta _{J,-}})\sin \alpha _{J,\mu }\cos \alpha _{J,\mu }, \\
t_{k,\mu }^{+-} &=&(\cos \eta _{J,+}e^{-i\eta _{J,+}}-\cos \eta
_{J,-}e^{-i\eta _{J,-}})\sin \alpha _{J,\mu }\cos \alpha _{J,\mu }, \\
r_{k,\mu }^{+-} &=&-i(\sin \eta _{J,+}e^{-i\eta _{J,+}}-\sin \eta
_{J,-}e^{-i\eta _{J,-}})\sin \alpha _{J,\mu }\cos \alpha _{J,\mu }.
\end{eqnarray*}
\end{small}
Here, the phase shifts are given as $\eta _{J,\pm }=\tan ^{-1}\frac{\Delta _{J,\pm }}{k}$, with the effective potentials
$\Delta_{J,+}=\frac{1}{2}(J\lambda _{0}+J\lambda )$,$\Delta_{J,-}=\frac{1}{2}\left[J\lambda _{0}-\left(J+1\right)\lambda \right]$. The Clebsch-Gordan coefficients $\cos\alpha_{J,\mu}$ and $\sin\alpha_{J,\mu}$ are given as
$\cos\alpha_{J,\mu}=\sqrt{\frac{J+\mu +\frac{1}{2}}{2J+1}}$,\quad$\sin \alpha_{J,\mu }=\sqrt{\frac{J-\mu +\frac{1}{2}}{2J+1}}$.
 
\section{Kraus Operators}
Now we express the Kraus operators $\mathcal{K}_{k,s;k',s'}$ in the basis set $\{|J,m\rangle\}$ based on the scattering matrix $\mathcal{S}$ obtained above. The block form of $\mathcal{S}$ means that $\mathcal{K}_{k,s;k',s'}$ is non-zero only if $k$ and $k'$ have the same absolute values. For example, we have
\begin{eqnarray}
&&\mathcal{K}_{k,+;k,+}\nonumber\\
&=&\langle k,+|\mathcal{S}|k,+\rangle\nonumber\\
&=&\left(
\begin{array}{ccccc}
t_{k,J+\frac{1}{2}} & \cdots& 0 & \cdots & 0  \\
\vdots & \ddots & 0 & \ddots & \vdots \\
0 & 0 & t_{k,m+\frac{1}{2}}^{++} & 0 & 0  \\
\vdots & \ddots & 0 & \ddots & \vdots \\
0 &  \cdots & 0 & \cdots &t_{k,-J+\frac{1}{2}}
\end{array}\right),\label{SKraus1}
\end{eqnarray}
which is a $(2J+1)$-dimension diagonal matrix. With the Clebsch-Gordan coefficients, the diagonal elements are rewritten as
\begin{eqnarray*}
t_{k,m+\frac{1}{2}}^{++}=(\xi+\frac{1}{2})+\zeta m,
\end{eqnarray*} 
where
\begin{small}
\begin{eqnarray*}
\xi&=&\frac{J+1}{2J+1}\cos\eta_{J,+}e^{-i\eta_{J,+}}+\frac{J}{2J+1}\cos\eta_{J,-}e^{-i\eta_{J,-}}-\frac{1}{2}\\
   &=&-i(\frac{J+1}{2J+1}\sin\eta_{J,+}e^{-i\eta_{J,+}}+\frac{J}{2J+1}\sin\eta_{J,-}e^{-i\eta_{J,-}})+\frac{1}{2},\\
\zeta&=&\frac{1}{2J+1}(\cos\eta_{J,+}e^{-i\eta_{J,+}}-\cos\eta_{J,-}e^{-i\eta_{J,-}})\\
     &=&-i\frac{1}{2J+1}(\sin\eta_{J,+}e^{-i\eta_{J,+}}-\sin\eta_{J,-}e^{-i\eta_{J,-}}).
\end{eqnarray*}
\end{small}
Considering the matrix form of the angular momentum operator $\widehat{\mathsf{J}}_{z}$ in the basis set $\{|J,m\rangle\}$, the matrix (\ref{SKraus1}) shows that the Kraus operator $\mathcal{K}_{k,+;k,+}$ is just
\begin{eqnarray*}
\mathcal{K}_{k,+;k,+}=(\xi+\frac{1}{2})\widehat{\mathsf{J}}_{0}+\zeta\widehat{\mathsf{J}}_{z},
\end{eqnarray*}
where $\widehat{\mathsf{J}}_{0}$ is the unit matrix. 

Similarly, the other Kraus operators are written in the compact form as  
\begin{eqnarray*}
\mathcal{K}_{k,s;\pm k,s}=(\xi\pm\frac{1}{2})\widehat{\mathsf{J}}_{0}+s\zeta\widehat{\mathsf{J}}_{z},\quad
\mathcal{K}_{k,-s;\pm k,s}=\zeta\widehat{\mathsf{J}}_{s}.
\end{eqnarray*}

\section{Quantum Master Equation}
The master equation (\ref{Master}) in the main text is obtained by substituting the Kraus operators (\ref{KrausOp}) into Eq.~(\ref{DynMap}) there. The calculations are straightforward, and yield the explicit expressions for the operators $\mathcal{T}_{0}$ and $\mathcal{T}$ as 
\begin{eqnarray*}
\mathcal{T}_{0} &\equiv &(|\xi |^{2}-
\frac{1}{4})\rho^{J}+|\zeta|^{2}(\widehat{\mathsf{J}}
_{z}\rho^{J}\widehat{\mathsf{J}}_{z}+\widehat{\mathsf{J}}_{+}\rho^{J}
\widehat{\mathsf{J}}_{-})+h.c., \\
\mathcal{T}_{x}&\equiv&2\xi\zeta^{\ast}\rho^{J}\widehat{\mathsf{J}}_{x}
+|\zeta|^{2}(\widehat{\mathsf{J}}_{z}\rho^{J}
\widehat{\mathsf{J}}_{+}-\widehat{\mathsf{J}}_{+}\rho^{J}\widehat{\mathsf{J}}_{z})+h.c., \\
\mathcal{T}_{y} &\equiv &2\xi\zeta^{\ast}\rho^{J}\widehat{\mathsf{J}}_{y}
+i|\zeta|^{2}(\widehat{\mathsf{J}}_{+}\rho^{J}\widehat{\mathsf{J}}_{z}-\widehat{\mathsf{J}}_{z}
\rho^{J}\widehat{\mathsf{J}}_{+})+h.c.,\\
\mathcal{T}_{z} &\equiv &2\xi\zeta^{\ast }\rho^{J}\widehat{\mathsf{J}}_{z}+|\zeta |^{2}
(\widehat{\mathsf{J}}_{+}\rho^{J}\widehat{\mathsf{J}}_{-}-
\widehat{\mathsf{J}}_{-}\rho^{J}\widehat{\mathsf{J}}_{+})+h.c..
\end{eqnarray*}
The terms containing a single angular momentum operators in $\mathcal{T}$ can be interpreted as a field-like torque, while the terms including two angular momentum operators the Slonczewski-type torque and the quantum fluctuations. This becomes clearer in the spin coherent state representation. 

\section{The Fokker-Planck Equation}
Here we explain the derivation of the Fokker-Planck equation (6) in the paper. In the spin coherent state representation $\{|J,\Omega\rangle\}$, the density matrix $\rho_{J}$ is expressed as\cite{Dalgebra}
\begin{eqnarray}
\rho^{J}=\int d\Omega\mathcal{P}_{J}(\Omega)|J,\Omega\rangle\langle J,\Omega|.\label{SPrho}
\end{eqnarray}
Using $\mathbf{S}=(\alpha,\beta,\gamma)$, and substituting the expression (\ref{SPrho}) into the master equation (5) in the paper, and utilizing the differential forms of the operators\cite{Dalgebra} $\widehat{\mathsf{J}}_{i}|J,\Omega\rangle\langle J,\Omega|\widehat{\mathsf{J}}_{j}$ ($i,j=0,+,-,z$), we derive the differential equation for $\mathcal{P}_{J}(\Omega)$ as
\begin{eqnarray}
\frac{\partial\mathcal{P}_{J}}{\partial t}&=&\frac{1}{\sin\Theta}\frac{\partial(-\mathsf{T}_{\Theta }\mathcal{P}_{J})}{\partial\Theta}+\frac{1}{\sin\Theta}\frac{\partial(-\mathsf{T}_{\Phi}\mathcal{P}_{J})}{\partial\Phi}\nonumber\\
&+&\frac{1}{\sin\Theta}\frac{\partial}{\partial\Theta}[\sin\Theta\frac{\partial(\mathcal{D}\mathcal{P}_{J})}{\partial\Theta}]
+\frac{1}{\sin^{2}\Theta}\frac{\partial^{2}(\mathcal{D}\mathcal{P}_{J})}{\partial\Phi ^{2}}.\nonumber\\ \label{SFPE}
\end{eqnarray}
Here,
\begin{eqnarray*}
\mathsf{T}_{\Theta}&=&\mathcal{A}(\alpha\cos\Theta\cos\Phi+\beta\cos\Theta\sin\Phi-\gamma\sin\Theta)\\
&+&\mathcal{B}(\alpha\sin\Phi-\beta\cos\Phi),\\
\mathsf{T}_{\Phi}&=&\mathcal{A}(-\alpha\sin\Phi+\beta\cos\Phi)\\
&+&\mathcal{B}(\alpha\cos\Theta\cos\Phi+\beta\cos\Theta\sin\Phi-\gamma\sin\Theta),\\
\mathcal{D}&=&\frac{\mathcal{A}}{2J+1}(1-\alpha\sin\Theta\cos\Phi-\beta\sin\Theta\sin\Phi-\gamma\cos\Theta),
\end{eqnarray*}
with the coefficients $\mathcal{A}=(2J+1)\frac{|\zeta|^{2}}{\tau},\mathcal{B}=\frac{2\Im(\xi^{\ast}\zeta)}{\tau}$.
Further analysis shows that $\mathsf{T}_{\Theta}$ and $\mathsf{T}_{\Phi}$ are the components of the spin transfer torque $\mathbf{T}=\mathcal{A}(\widehat{\mathbf{m}}\times\mathbf{S})\times\widehat{\mathbf{m}}
+\mathcal{B}\widehat{\mathbf{m}}\times\mathbf{S}$ in the spherical coordinates, where the unit vector
$\widehat{\mathbf{m}}$ denotes the direction of the macrospin. Replacement of the differential operators in spherical coordinates by the divergence operator $\nabla$ and Laplace operator $\nabla^{2}$ reduces Eq.~(\ref{SFPE}) to the simple form of the Fokker-Planck equation (FPE) 
\begin{eqnarray}
\frac{\partial}{\partial t}\mathcal{P}_{J}(\widehat{\mathbf{m}},t)
=-\nabla\cdot(\mathbf{T}\mathcal{P}_{J})+\nabla^{2}(\mathcal{D}\mathcal{P}_{J}). \label{SFPE2}
\end{eqnarray}


\begin{thebibliography}{99}
\bibitem{OpenQuan1} H. J. Carmichael, \emph{Statistical methods in quantum optics 1:
Master equations and Fokker-Planck equations} (Springer,
Berlin, 1999).
\bibitem{OpenQuan1a} H. J. Carmichael, \emph{Statistical methods in quantum optics 2:
Nonclassical fields} (Springer, Berlin, 2008).
\bibitem{OpenQuan2} H.-P. Breuer and F. Petruccione, \emph{The Theory of Open
Quantum Systems} (Oxford University Press, 2002).
\bibitem{arecchi72} F. T. Arecchi, E. Courtens, R. Gilmore, and H. Thomas,
Phys. Rev. A \textbf{6}, 2211 (1972).
\bibitem{weedbrook12} C. Weedbrook, S. Pirandola, R. Garc´ıa-Patr´on, N. J. Cerf,
T. C. Ralph, J. H. Shapiro, and S. Lloyd, Rev. Mod. Phys.
\textbf{84}, 621 (2012).
\bibitem{aharoni96} A. Aharoni, \emph{Introduction to the Theory of Ferromagnetism}
(Clarendon Press, Oxford, 1996).
\bibitem{suhl07} H. Suhl, \emph{Relaxation processes in micromagnetics} (Oxford
University Press, 2007).
\bibitem{revtorq1} D. Ralph and M. Stiles, Journal of Magnetism and Magnetic
Materials \textbf{320}, 1190 (2008).
\bibitem{revtorq2} A. Brataas, A. D. Kent, and H. Ohno, Nat. Mater \textbf{11}, 372
(2012).
\bibitem{LLG} J. Z. Sun, Phys. Rev. B \textbf{62}, 570 (2000).
\bibitem{wernsdorfer07} W. Wernsdorfer, Advances in Chemical Physics \textbf{118}, 99
(2007).
\bibitem{torq1} J. C. Slonczewski, Journal of Magnetism and Magnetic Materials
\textbf{159}, L1 (1996).
\bibitem{torq2} L. Berger, Phys. Rev. B \textbf{54}, 9353 (1996).
\bibitem{Magnon} M. Madami, S. Bonetti, G. Consolo, S. Tacchi, G. Carlotti,
G. Gubbiotti, F. B. Mancoff, M. A. Yar, and J. \r{A}kerman,
Nature Nanotech. \textbf{6}, 635 (2011).
\bibitem{MonCar} Y. Wang and L. J. Sham, Phys. Rev. B \textbf{85}, 092403 (2012).
\bibitem{Kraus} K. Krause, \emph{States, Effects, And Operations Fundamental
Notions Of Quantum Theory}, edited by A. B¨ohm, J. Dollard,
and W. Wootters (Springer-Verlag, Berlin, 1983).
\bibitem{Dalgebra} L. M. Narducci, C. M. Bowden, V. Bluemel, G. P. Garrazana,
and R. A. Tuft, Phys. Rev. A \textbf{11}, 973 (1975).
\bibitem{CurrFluc} V. E. Demidov, S. Urazhdin, E. R. J. Edwards, M. D.
Stiles, R. D. McMichael, and S. O. Demokritov, Phys.
Rev. Lett. \textbf{107}, 107204 (2011).
\bibitem{Brown} W. F. Brown, Phys. Rev. \textbf{130}, 1677 (1963).
\bibitem{Therm} Z. Li and S. Zhang, Phys. Rev. B \textbf{69}, 134416 (2004).
\bibitem{WKB} H. Risken, \emph{The Fokker-Planck Equation} (Springer-Verlag,
Berlin, 1984).
\bibitem{Method} R. Courant and D. Hilbert, \emph{Methods of Mathematical
Physics} (Interscience, New York, 1953).
\bibitem{SemiScat} J. Foros, A. Brataas, Y. Tserkovnyak, and G. E.W. Bauer,
Phys. Rev. Lett. \textbf{95}, 016601 (2005).
\bibitem{Tbit} J. K. W. Yang, Y. Chen, T. Huang, H. Duan, N. Thiyagarajah,
H. K. Hui, S. H. Leong, and V. Ng, Nanotechnology
\textbf{22}, 385301 (2011).
\end{thebibliography}
\end{document}